%% file: main.tex
\documentclass[sigconf]{acmart}
\AtBeginDocument{%
  }

\setcopyright{acmlicensed}
\copyrightyear{2018}
\acmYear{2018}
\acmDOI{XXXXXXX.XXXXXXX}
\acmConference[Conference acronym 'XX]{Make sure to enter the correct
  conference title from your rights confirmation email}{June 03--05,
  2018}{Woodstock, NY}
\acmISBN{978-1-4503-XXXX-X/2018/06}

\usepackage{booktabs} 
\usepackage{multirow} 
\usepackage{tabularx}
\usepackage{caption}
\usepackage{enumitem}
\usepackage{subcaption}
\usepackage{caption}
\usepackage{subcaption} 

\usepackage{booktabs}
\usepackage{multirow}
\usepackage{colortbl}
\usepackage{siunitx}
\usepackage{graphicx}
\usepackage{booktabs}
\usepackage{amsmath}
        
\usepackage{amssymb}
\usepackage{amsfonts}

\begin{document}

\title{Taming the Long Tail: Denoising Collaborative Information for Robust Semantic ID Generation}

\author{Yi Xu}
\affiliation{
  \institution{Alibaba International Digital Commerce Group}
  \city{Beijing}\country{China}
}
\email{xy397404@alibaba-inc.com}

\author{Moyu Zhang}
\affiliation{
  \institution{Alibaba International Digital Commerce Group}
  \city{Beijing}\country{China}
}
\email{zhangmoyu.zmy@alibaba-inc.com}

\author{Chaofan Fan}
\affiliation{
  \institution{Alibaba International Digital Commerce Group}
  \city{Beijing}\country{China}
}
\email{fanchaofan.fcf@alibaba-inc.com}

\author{Jinxin Hu}
\affiliation{
  \institution{Alibaba International Digital Commerce Group}
  \city{Beijing}\country{China}
}
\email{jinxin.hjx@alibaba-inc.com}
\authornote{Corresponding author}

\author{Yu Zhang}
\affiliation{
  \institution{Alibaba International Digital Commerce Group}
  \city{Beijing}\country{China}
}
\email{daoji@alibaba-inc.com}

\author{Xiaoyi Zeng}
\affiliation{
  \institution{Alibaba International Digital Commerce Group}
  \city{Beijing}\country{China}
}
\email{yuanhan@taobao.com}

\author{Jing Zhang}
\affiliation{
  \institution{Wuhan University, School of Computer Science}
  \city{Wuhan}\country{China}
}
\email{jingzhang.cv@gmail.com}

\renewcommand{\shortauthors}{Trovato et al.}

\begin{abstract}
Item IDs form the backbone of industrial recommender systems, but suffer from representation instability and poor long-tail generalization in large, dynamic item corpora. Semantic IDs (SIDs) mitigate these issues by enabling knowledge sharing through quantization of item content features. Existing methods attempt to enhance SID expressiveness by incorporating collaborative information with content features; however, they often overlook a critical distinction: unlike relatively uniform content features, user-item interactions are highly skewed, resulting in a significant quality gap in collaborative information between popular and long-tail items. This mismatch leads to two critical limitations:
(1) Collaborative Noise Corrupts Behavior–Content Alignment: Behavior–content alignment has become a prevailing approach for modeling behavior-content shared information. However, indiscriminate alignment allows collaborative noise from long-tail items to corrupt their content representations, leading to the loss of critical multimodal information.
(2) Collaborative Noise Obscures Critical Behavioral SIDs: When modeling modality-specific information, prior works typically generate multiple behavioral SIDs with equal weights for each item. This equal-weight scheme fails to reflect the varying importance of different behavioral SIDs, making it difficult for downstream tasks to distinguish informative SIDs from noisy ones. 
To address these challenges, we propose \textbf{ADC-SID}, a framework that \textbf{A}daptively \textbf{D}enoises \textbf{C}ollaborative information for SID quantization. It comprises two key components:
(i) Adaptive Behavior–Content Alignment, which adaptively adjusts the alignment strength to mitigate the corruption caused by collaborative noise;
(ii) Dynamic Behavioral Weighting Mechanism, which learns importance scores for behavioral SIDs to enable downstream recommendation models to suppress noisy SIDs.
Extensive experiments on both public and industrial datasets demonstrate ADC-SID’s superiority in generative retrieval and discriminative ranking tasks.

\end{abstract}





\maketitle
\section{Introduction}
\input{secs/intro_v5}
\section{Related Work}
\input{secs/related_works_v1}

\section{Methodology}
\input{secs/method_v1}
\section{Experiments}
\input{secs/exp_v3}
\section{Online Experiments}
We conducted a 5-day online A/B test on both the generative retrieval system and the discriminative ranking system of a large-scale e-commerce platform to evaluate the effectiveness of the proposed method. The experimental group, employing our designed 8-token SID, was allocated 10\% of random user traffic, in comparison with the production Item ID-based system. The experimental results show that ADC-SID achieved significant improvements in key business metrics: in the generative retrieval task, Advertising Revenue increased by 3.50\% and Click-Through Rate (CTR) increased by 1.15\%; in the discriminative ranking task, Advertising Revenue increased by 1.56\% and CTR increased by 3.04\%. These online experience results fully demonstrate the practical value and production-readiness of our proposed method.

\begin{table}[ht]
\centering
\caption{Online A/B Test results in two scenarios: Generative Retrieval and Discriminative Ranking.}
\label{tab:online_scenarios}
\begin{tabular}{l|cc|cc}
\toprule
\textbf{Scenario} & \multicolumn{2}{c|}{\textbf{Generative Retrieval}} & \multicolumn{2}{c}{\textbf{Discriminative Ranking}} \\
\cmidrule(lr){2-3} \cmidrule(lr){4-5}
\textbf{Metric} & \textbf{Revenue} & \textbf{CTR} & \textbf{Revenue} & \textbf{CTR} \\
\midrule
ADC-SID & +3.50\% & +1.15\% & +1.56\% & +3.04\% \\
\bottomrule
\end{tabular}
\end{table}

\section{Conclusion}
In this work,  we propose ADC-SID, a novel framework designed to denoise collaborative information at the stages of behavioral-content alignment and behavior-specific modeling. By effectively filtering out collaborative noise from long-tail items, ADC-SID enables adaptive multimodal fusion and maximizes the preservation of informative modal information. Our approach integrates two key innovations: (i) Adaptive Behavior–Content Alignment, which adaptively adjusts the alignment strength to mitigate the corruption caused by collaborative noise;
(ii) Dynamic Behavioral Weighting Mechanism, which learns importance scores for behavioral SIDs to enable downstream recommendation models to suppress noisy SIDs.

Extensive offline experiments and large-scale online A/B experiments demonstrate that ADC-SID consistently improves the prediction accuracy of both generative retrieval and discriminative ranking across diverse datasets. 
In future research, we plan to extend these principles to the user side, aiming to further improve personalization by jointly denoising and weighting user representations. Additionally, we will explore more efficient model architectures to reduce the computational overhead of multi-expert systems, facilitating broader deployment in large-scale, latency-sensitive production environments.

\bibliographystyle{ACM-Reference-Format}
\bibliography{main}

\end{document}

%% file: secs/intro_v5.tex
Recommender systems commonly represent items using unique item IDs. 
While existing item ID-based models excel at representing popular items, they struggle with the inherent long-tail distribution of item popularity. A vast majority of items suffer from severe data sparsity, which prevents the model from learning robust and reliable representations, ultimately degrading overall predictive performance\cite{rajput2023recommendersystemsgenerativeretrieval, zheng2025enhancingembeddingrepresentationstability,kudo2018subwordregularizationimprovingneural}.

\begin{figure*}[t]
\centering
\includegraphics[width=\textwidth]{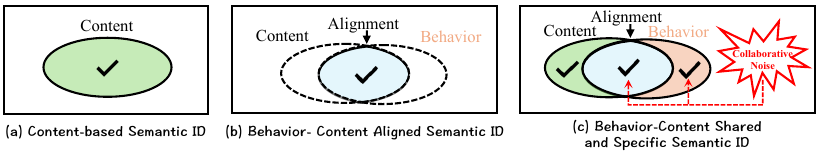}
\centering
\caption{
Illustration of SIDs Quantization Paradigm.(a) Content-based SIDs: Quantize content features (e.g., title, images) of items into SIDs.
(b) Behavior-Content Aligned SIDs: Capture the behavior–content shared information through alignment for SIDs.
(c) Behavior-content Shared and Specific SIDs: Learn both behavior–content shared and specific information for SIDs.
}

\label{fig:intro1}
\end{figure*}

Content-based SIDs alleviate the long-tail item problem by quantizing item content (e.g., title, images), allowing similar items to share identifiers\cite{rajput2023recommendersystemsgenerativeretrieval,hou2023learningvectorquantizeditemrepresentation}, as shown in Fig. \ref{fig:intro1}~(a). 
Considering there exists a gap between the content domain and the behavior domain: user–item interactions induce dynamic behavioral properties (e.g., popularity, associated user groups) that content alone cannot capture, creating a performance ceiling for content‑based SIDs\cite{ye2025dasdualalignedsemanticids,wang2024eagertwostreamgenerativerecommender}.
To narrow this gap, recent methods incorporate collaborative information into SID quantization\cite{tan2024idgenrecllmrecsysalignmenttextual,zheng2024adaptinglargelanguagemodels,Wang_2024}.
A common strategy is behavior–content alignment, which aligns content representations with pre-trained item embeddings learned from user–item interactions\cite{wang2024eagertwostreamgenerativerecommender,wang2025empoweringlargelanguagemodel}. 
These methods successfully capture behavior–content shared information, as shown in Fig. \ref{fig:intro1}~(b).
However, they often overlook modality-specific characteristics, especially in the behavioral modality, which is tightly coupled with user interactions. 
Recently, modeling both the modality-shared and modality-specific characteristics has made progress in multimodal SID quantizations\cite{xu2025mmq}, as shown in Fig. \ref{fig:intro1}~(c).
\begin{figure}[t]
\centering
\includegraphics[width=0.9\columnwidth]{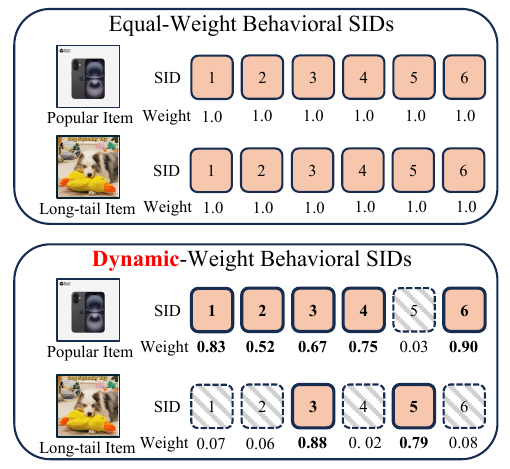}
\caption{
Equal-Weight SID paradigm and the Proposed Dynamic-Weight SID paradigm for Behavioral SIDs
}
\vspace{-0.3cm}
\label{fig:intro2}
\end{figure}

Despite significant progress in the above semantic ID generation works, these methods still fail to adequately address the inherent modality gap between collaborative information and content. Specifically, the collaborative information derived from user-item interactions suffers from extreme sparsity and a long-tail distribution. This, in turn, creates a significant quality gap in the learned representations of popular versus long-tail items\cite{yan_learning_2021,zhao_autoemb_2020,Joglekar2019NeuralIS,Sun2020AGN}. This fundamental mismatch between the behavior and content modalities presents two critical challenges for existing methods:

(1) Collaborative Noise Corrupts Behavior–Content Alignment:
In modeling the shared information between behavioral and content modalities, prior methods have often employed an indiscriminate alignment approach. This strategy causes collaborative noise, stemming from the sparse and unreliable interaction signals of long-tail items, to corrupt the otherwise robust content representations. The resulting noisy SIDs lose their effectiveness, which in turn degrades the performance of downstream recommendation tasks. Therefore, how to effectively minimize the impact of this collaborative noise on content representations during the alignment process, particularly for long-tail items, is a crucial but unresolved challenge.

(2) Collaborative Noise Obscures Critical Behavioral SID:
While current methods generate multiple SIDs to preserve rich modal-specific information, they typically adopt an equally-weighted SID paradigm\cite{wang2025empoweringlargelanguagemodel,wang2024eagertwostreamgenerativerecommender}, as shown in Fig.~\ref{fig:intro2}. This uniform weighting scheme may be adequate for content features, which are often evenly distributed across items. However, it fundamentally fails to account for the vast disparity in the quality and density of collaborative information between popular and long-tail items. For popular items, the diverse user interactions can be effectively captured by multiple behavioral SIDs. For long-tail items, in contrast, the sparse interactions mean that the generated behavioral SIDs are inherently noisy. It is likely that only a small subset of these SIDs of a long-tail item captures meaningful collaborative information, while the majority constitute noise. When all SIDs are weighted equally, the information from the few effective SIDs is easily overwhelmed by the cumulative noise from the many unreliable ones for a long-tail item. This poor information-to-noise ratio degrades the quality of the final long-tail item representation and, consequently, harms the accuracy of downstream recommendations. Therefore, developing a mechanism to denoise the behavioral SIDs for long-tail items also represents a critical, yet unaddressed, challenge in the SID field.

\begin{figure*}[t]
\includegraphics[width=0.77\textwidth]{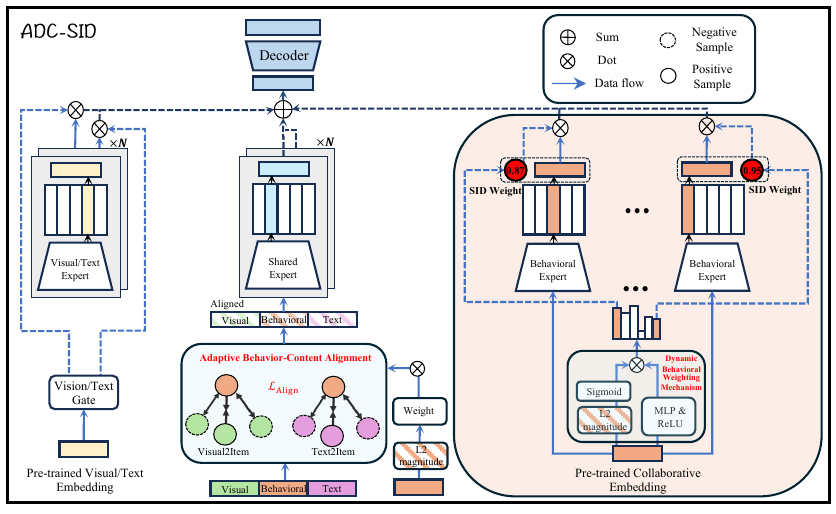}
\centering
\caption{
The overall architecture of ADC-SID. This framework
comprises two key components: (i) Adaptive Behavior–Content Alignment, which adaptively adjusts the alignment strength to mitigate the corruption caused by collaborative noise; (ii) Dynamic Behavioral Weighting Mechanism, which learns importance scores for behavioral SIDs to enable downstream recommendation models to suppress noisy SIDs.
}
\label{fig:realexp}
\end{figure*}

To address the above two challenges, we propose ADC-SID, a novel framework designed to denoise collaborative information at the stages of behavioral-content alignment and behavior-specific modeling. By effectively filtering out collaborative noise from long-tail items, ADC-SID enables adaptive multimodal fusion and maximizes the preservation of informative modal information.
This framework comprises two key components: (1) Adaptive Behavior-Content Alignment: We introduce an alignment controller that adjusts the alignment strength based on the reliability of pre-trained item representations, which migrates the collaborative noise for alignment.(2) Dynamic Behavioral Weighting Mechanism: We propose a dynamic behavioral weighting mechanism that learns importance scores for behavioral SIDs, enabling the downstream recommendation to suppress noisy SIDs and improve downstream recommendation performance. The overall framework of ADC-SID is shown in Fig. \ref{fig:realexp}

Our contributions are as follows:
\begin{itemize}
    \item To the best of our knowledge, we are the first to adaptively denoise collaborative signals in SID quantization to robustly learn behavior-content shared and specific information, thereby enhancing the expressiveness of SIDs and improving accuracy in downstream recommendation tasks.
    \item We propose an adaptive behavior-content alignment that adaptively adjusts the alignment strength, mitigating collaborative noise from long-tail items. 
    \item We propose a dynamic behavioral weighting mechanism that learns importance scores for behavioral SIDs, effectively suppressing noisy SIDs in downstream recommendation.
    \item Extensive offline experiments and online A/B tests across both generative and discriminative recommendation tasks demonstrate the effectiveness and robustness of our approach.
\end{itemize}

%% file: secs/related_works_v1.tex
Traditional ItemIDs limit the generalization of recommender systems due to their lack of semantics\cite{10.1145/2523813,jin2024languagemodelssemanticindexers,luo2024qarmquantitativealignmentmultimodal, kuai2024breakinghourglassphenomenonresidual, zheng2025pretraininggenerativerecommendermultiidentifier,yang2025sparsemeetsdenseunified}. To address this, content-based SIDs quantize item content features into discrete codes. TIGER\cite{rajput2023recommendersystemsgenerativeretrieval} pioneered this with RQ-VAE\cite{zhu2024scalingcodebooksizevqgan}, while later works like SPM-SID \cite{singh2024bettergeneralizationsemanticids} and PMA \cite{zheng2025enhancingembeddingrepresentationstability}, inspired by Large Language Models, explored more granular subword-based compositions\cite{kudo2018subwordregularizationimprovingneural}. 
To incorporate dynamic behavioral information into SIDs, recent work can be broadly categorized into two approaches. The first approach involves injecting explicit collaborative signals. For instance, LC-Rec\cite{zheng2024adaptinglargelanguagemodels} designs a series of alignment tasks to unify semantic and collaborative information. ColaRec\cite{Wang_2024} distills collaborative signals directly from a pre-trained recommendation model and combines them with content information. IDGenRec\cite{tan2024idgenrecllmrecsysalignmenttextual}  leverages LLMs to generate semantically rich textual identifiers, showing strong potential in zero-shot settings. The second approach focuses on aligning pre-trained representations. To this end, recent methods introduce pre-trained collaborative representations and align content representations with them. For example, EAGER\cite{wang2024eagertwostreamgenerativerecommender} generates separate collaborative and content SIDs using K-means on pre-trained embeddings and then aligns them in downstream tasks. DAS\cite{ye2025dasdualalignedsemanticids} employs multi-view contrastive learning to maximize the mutual information between SIDs and collaborative signals. LETTER\cite{wang2024learnableitemtokenizationgenerative} integrates hierarchical semantics, collaborative signals and code assignment diversity to generate behavior-content fused SIDs with RQ-VAE. MM-RQ-VAE\cite{wang2025empoweringlargelanguagemodel} generates collaborative SIDs, textual SIDs and visual SIDs with pre-trained collaborative embeddings and multimodal embeddings, and introduces contrastive learning for behavior-content alignment.

%% file: secs/method_v1.tex
This section details the proposed ADC-SID framework, which is proposed to adaptively denoise collaborative signals for SID quantization. We introduce the SID quantization formulation in Section \ref{sec:section3_1} and the overall architecture of ADC-SID in Section \ref{sec:section3_2}. The two key components of ADC-SID are introduced in Sections \ref{sec:section_3_3} and \ref{sec:section3_4}:
(1) Adaptive Behavior-Content Alignment: To mitigate the collaborative noise from long-tail items that corrupts behavior–content alignment, we propose an alignment strength controller to dynamically adjust the behavior–content alignment strength based on the information richness of pre-trained item representations, detailed in Section \ref{sec:section_3_3}. (2) Dynamic Behavioral Weighting Mechanism: To address the challenge that crucial signals are easily overwhelmed by noisy SIDs under the equal-weight SID paradigm, we propose a dynamic behavioral weighting mechanism that adaptively learns importance scores for behavioral SIDs, enabling the downstream recommendation to suppress uninformative SIDs and improve downstream recommendation performance, detailed in Section \ref{sec:section3_4}.
\subsection{Problem Formulation}
\label{sec:section3_1}
The item tokenizer is designed to quantize the pre-trained textual, visual, and collaborative embeddings of items into a sequence of discrete SIDs.
Formally, for a given item, we first leverage pre-trained vision and text embedding models to obtain items pre-trained vision embedding $\mathbf{e}_{v}$ and pre-trained text embedding $\mathbf{e}_{t}$. 
The pre-trained collaborative embedding $\mathbf{e}_{b}$ is obtained from SASRec\cite{kang2018selfattentivesequentialrecommendation} trained with user-item interactions.
The item tokenizer $\mathcal{T}_{\text{item}}$ then quantizes the pre-trained collaborative embeddings into a sequence of discrete SIDs. The formulation is as follows.
\begin{equation}
\label{eq:item_tokenization} 
  Semantic\_IDs = (c_1, c_2, \dots, c_l) = \mathcal{T}_{\text{item}}([\mathbf{e}_{t},\mathbf{e}_{v},\mathbf{e}_{b}]) 
\end{equation}
where $l$ is the length of the SIDs, $c_i$ is the $i$-th SID. 

\subsection{Behavior-Content Mixture-of-Quantization Network}
\label{sec:section3_2}
To simultaneously capture both behavior-content shared and specific information, we propose the behavior-content mixture-of-quantization network, where shared experts learn shared information across behavior and content modalities, and specific experts focus on modality-specific information.
\subsubsection{Behavior-Content Shared Experts}
The shared experts are designed to quantize the aligned behavior-content representation into shared latent embeddings, which are used to generate modality-shared SIDs.
For a given item, its pre-trained textual, visual, and behavioral embeddings are first projected into a high-dimensional space via shallow two-layer neural networks, denoted $D_t,D_v,D_b$ respectively. 
The hidden representations are denoted as $\mathbf{h_t},\mathbf{h_v},\mathbf{h_b}$. 
\begin{equation}
\label{eq:shared_hidden_representation} 
\mathbf{h_t} = D_t(\mathbf{e_t}),\mathbf{h_v} = D_v(\mathbf{e_v}),\mathbf{h_b} = D_b(\mathbf{e_b})\\
\end{equation}
\begin{equation}
\mathbf{h}=[\mathbf{h_t},\mathbf{h_v},\mathbf{h_b}]
\end{equation}
To learn the aligned behavior-content information, these projected hidden representations are optimized by the adaptive behavior-content alignment mechanism, detailed in Section~\ref{sec:section_3_3}. For the $i$-th shared expert $E_{s,i}$, the hidden representations $\mathbf{h}$ are encoded into a shared latent embedding $\mathbf{z}_{s,i}$, as formatted in Eq.\ref{eq:share_encode}. The shared latent embedding $\mathbf{z}_{s,i}$ is used to search for the most similar codeword from the shared codebook $C_{s,i} = \{ \mathbf{z_{q_{s,j}}} \}_{k=1}^K$, where $K$ indicates the codebook size, and $i \in \{1, \dots, N_s\}$, $N_s$ denotes the number of shared experts. The most similar codeword index $c_{s,i}$ is searched by maximizing the cosine distance between $\mathbf{z}_{s,i}$ and all codewords in $C_{s,i}$, as formatted in Eq.\ref{eq:cosine}.
\begin{gather}
\label{eq:share_encode} 
\mathbf z_{s,i}= E_{s,i}(\mathbf{h})\\
\label{eq:cosine} 
    c_{s,i}
=\underset{j\in\{1,\dots,K\}}{\arg\max}\;
\frac{\mathbf z_{s,i}^\top \mathbf z_{q_{s,j}}}
{\lVert \mathbf z_{s,i}\rVert\,\lVert \mathbf z_{q_{s,j}}\rVert}
\end{gather}
\subsubsection{Specific Experts of Behavior and Content Modalities}
The specific experts are designed to learn specific information about each modality and generate modality-specific SIDs. For each modality, there is a group of modality-specific experts and corresponding modality-specific codebooks. For example, for the textual modality, a set of specific experts $\{E_{t,i}\}_{i=1}^{N_t}$, transforms the original pretrained embedding $\mathbf{e}_t$ into a corresponding set of latent embeddings $\{\mathbf{z}_{t,i}\}_{i=1}^{N_t}$, as formatted in Eq.\ref{eq:specific_encode}. The textual SIDs $\{c_{t,i}\}_{i=1}^{N_t}$ are searched from codebooks $\{C_{t,i}\}_{i=1}^{N_t}$ with cosine distance, as formatted in Eq.\ref{eq:specific_cos}. 
Analogously, the latent embeddings of visual and behavioral modality are denoted as  $\{\mathbf{z}_{v,i}\}_{i=1}^{N_v}$, $\{\mathbf{z}_{b,i}\}_{i=1}^{N_b}$. The visual and behavior SIDs are denoted $\{c_{v,i}\}_{i=1}^{N_v}$, $\{c_{b,i}\}_{i=1}^{N_b}$. $N_v$, $N_b$ are the number of visual and behavioral experts separately.
 \begin{align}
&  \label{eq:specific_encode}
     \mathbf{z}_{t,i}=E_{t,i}(\mathbf{e}_t),\mathbf{z}_{v,i}=E_{v,i}(\mathbf{e}_v),\mathbf{z}_{b,i}=E_{t,i}(\mathbf{e}_b)\\
& \label{eq:specific_cos}
c_{t,i}
=\underset{j\in\{1,\dots,K\}}{\arg\max}\;
\frac{\mathbf z_{t,i}^\top \mathbf z_{q_{t,j}}}
{\lVert \mathbf z_{t,i}\rVert\,\lVert \mathbf z_{q_{t,j}}\rVert},\\
& c_{v,i}
=\underset{j\in\{1,\dots,K\}}{\arg\max}\;
\frac{\mathbf z_{s,i}^\top \mathbf z_{q_{v,j}}}
{\lVert \mathbf z_{s,i}\rVert\,\lVert \mathbf z_{q_{v,j}}\rVert},\\
& c_{b,i}
=\underset{j\in\{1,\dots,K\}}{\arg\max}\;
\frac{\mathbf z_{s,i}^\top \mathbf z_{q_{b,j}}}
{\lVert \mathbf z_{s,i}\rVert\,\lVert \mathbf z_{q_{b,j}}\rVert} 
 \end{align}
To better integrate the shared behavior-content information with modality-specific characteristics, we learn fusion weights for each modality via a gating mechanism. The weighted fusion is formulated as Eq.\ref{eq:fusion} and Eq.\ref{eq:fusion2}, where $R(\mathbf{e_b})_i$ is learned from the Dynamic Behavioral Weighting Mechanism, detailed in Section \ref{sec:section3_4}, $g_t$ and $g_v$ are gating mechanisms as formulated in Eq.\ref{eq:g_t}, Eq.\ref{eq:g_v}.

 \begin{gather}
 \label{eq:fusion}
     \mathbf{z} = \sum_{i=1}^{N_s} \mathbf{z}_{s,i} + \sum_{i=1}^{N_v}g_{v,i}\mathbf{z}_{v,i} + \sum_{i=1}^{N_t}g_{t,i}\mathbf{z}_{t,i}+\sum_{i=1}^{N_b} R(\mathbf{e_b})_i \mathbf{z}_{b,i}\\
     \label{eq:fusion2}
     \mathbf{z_q} = \sum_{i=1}^{N_s} \mathbf{z}_{q_{s,i}} + \sum_{i=1}^{N_v}g_{v,i}\mathbf{z}_{q_{v,i}} + \sum_{i=1}^{N_t}g_{t,i}\mathbf{z}_{q_{t,i}}+\sum_{i=1}^{N_b} R(\mathbf{e_b})_i \mathbf{z}_{q_{b,i}}\\
     \label{eq:g_t}
     g_{t}=softmax(MLP_t(\mathbf{e}_{t})+b_t)\\
      \label{eq:g_v}
     g_{v}=softmax(MLP_v(\mathbf{e}_{v})+b_v)\\    
     \label{eq:recon_loss}
     \mathcal{L}_{recon}=||\mathbf{e}-decoder(\mathbf{z}+sg(\mathbf{z}_q-\mathbf{z}))||^2
 \end{gather}
The decoder then reconstructs from the fused latent representations and codeword representations to the fused pre-trained embeddings $\mathbf{e=[\mathbf{e_t},\mathbf{e_v},\mathbf{e_b}]}$.  The reconstruction loss is formulated in Eq.\ref{eq:recon_loss}, where $sg(\cdot)$ denotes the stop-gradient operation.
\subsection{Adaptive Behavior-Content Alignment}
\label{sec:section_3_3}
To mitigate the collaborative noise from long-tail items that corrupts behavior–content alignment, 
when modeling the shared information between behavior and content, the alignment strength controller is proposed to dynamically adjust the behavior–content alignment strength based on the information richness of the pre-trained item representations.
\subsubsection{Alignment Strength Controller}
To adaptively align behavior and content information for head and long-tail items, the alignment strength controller is proposed to calculate the alignment strength. 
\begin{figure}[t]
\includegraphics[width=0.28\textwidth]{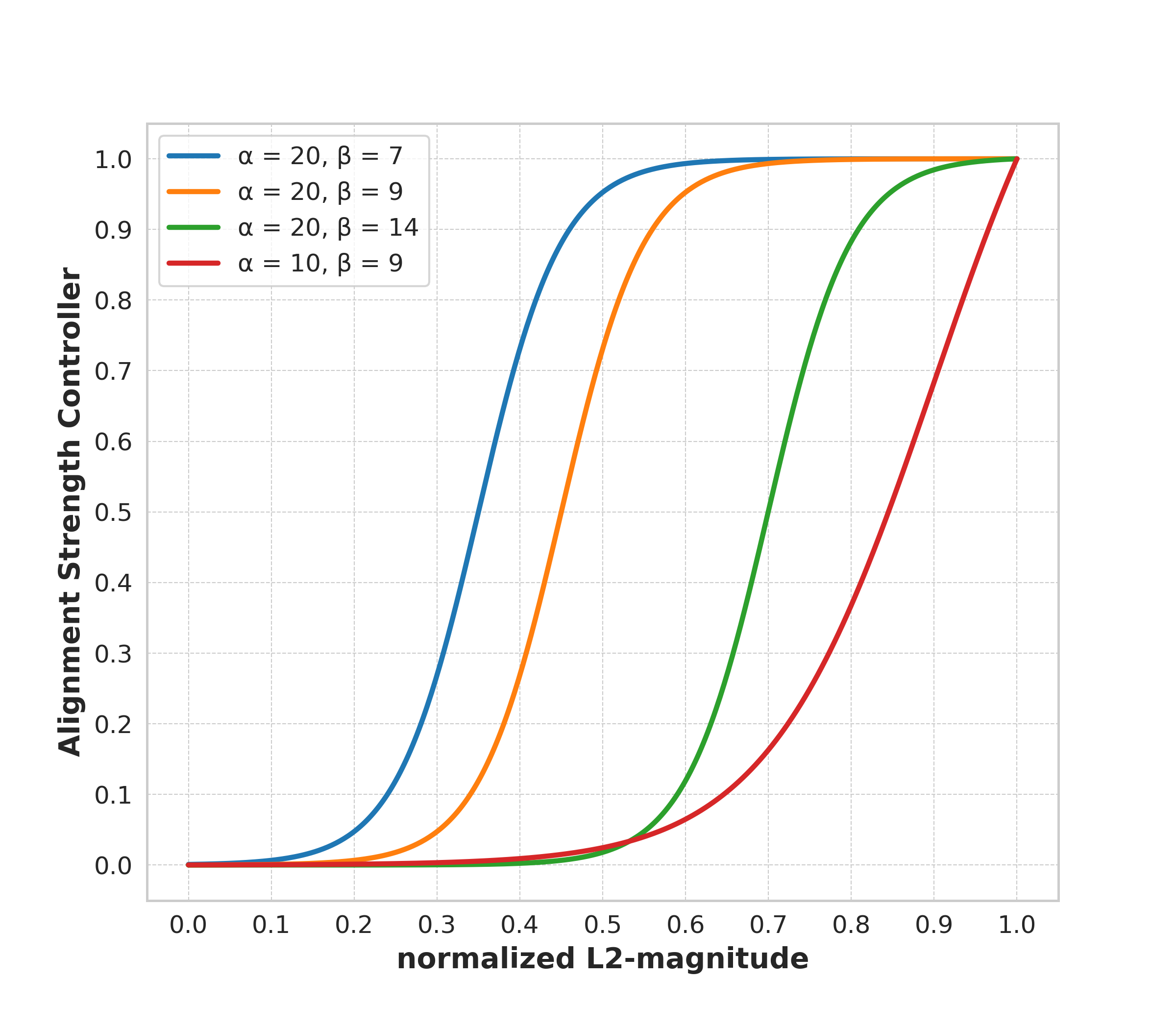}
\centering
\caption{
Illustration of alignment strength controller with different hyperparameters ($\alpha$, $\beta$).
}
\label{fig:controller}
\end{figure}
Long-tail items suffer from extreme user-item interaction sparsity, leading to the noisy pre-trained collaborative representations. In contrast, the pre-trained text and visual embeddings of long-tail items are reliable.
To mitigate the collaborative noise from long-tail items, the alignment strength of content information to behavioral information should be minimized. 
By contrast, popular items benefit from abundant user-item interactions, possessing representations rich in reliable collaborative information. Aligning the content information of head items to reliable collaborative information can effectively mitigate the gap between the content and behavioral domains, thereby enhancing the representational ability of SIDs. 

Given the long-tailed nature of user-item interactions in recommendation, L2-magnitudes of embeddings are indicative of training saturation: embeddings with abundant interactions (e.g., head items) typically achieve larger L2-magnitudes than those from sparse interactions. Based on these considerations, we use the L2-magnitude of an item's behavioral embedding as a proxy for its information richness.
\begin{gather}
\label{eq:controller1}
 N_{\max} = \max_{i \in \{1, \ldots, M\}} \left( \|\mathbf{e}_{b,i}\|_2 \right),N_{\min} = \min_{i \in \{1, \ldots, M\}} \left( \|\mathbf{e}_{b,i}\|_2 \right)\\
\label{eq:controller2}
 N_{\text{norm}}(\mathbf{e}_{b,j}) = \frac{\|\mathbf{e}_{b,j}\|_2 - N_{\min}}{N_{\max} - N_{\min}}\\
 \label{eq:controller3}
w=\frac{\sigma(\alpha N_{\text{norm}}(\mathbf{e}_{b,j})-\beta)}{\sigma(\alpha-\beta)}
\end{gather}
Specifically, consider the pre-trained collaborative embedding matrix $\mathbf{E} \in \mathbb{R}^{M \times D}$, consisting of $M$ vectors $\{\mathbf{e}_{b,1}, \mathbf{e}_{b,2}, \ldots, \mathbf{e}_{b,M}\}$, $M$ is the number of items. For the $j$-th embedding $\mathbf{e}_{b,j}$, the alignment strength controller is formulated as Eq.~\ref{eq:controller3}. $\alpha$ and $\beta$ are hyperparameters that jointly determine the steepness of the curve and the threshold for distinguishing long-tail items. By tuning $\alpha$ and  $\beta$ , the function can adapt to different data distributions. In our experiments, the optimal setting was found to be $\alpha=10$ and $\beta=9$. The illustration of alignment strength controller with different hyperparameters ($\alpha$, $\beta$) is shown in Fig.~\ref{fig:controller}.
\subsubsection{Behavior-Content Contrastive Learning}
To learn the shared information between behavior and content, the adaptive behavior-content contrastive learning is employed for $\langle Collaborative, Text \rangle$ and $\langle Collaborative, Visual \rangle$. Specifically, for the alignment between behavior and text modality, the adaptive contrastive learning is performed between the text representation $\mathbf{h_t}$ and the behavioral representation $\mathbf{h_b}$ to maximize the mutual information between the content modality and the behavior modality. Specifically, for a given item, its text representation and behavioral representation form a positive pair $\langle \mathbf{h_t}, \mathbf{h_{b^{+}}}\rangle$. The collaborative representations of all other items in the batch form negative pairs $\langle \mathbf{h_t}, \mathbf{h_{b_{i}^{-}}}\rangle$, where $i \in \{1, \dots, B\}$ and $B$ is the batch size. The contrastive learning of $\langle Collaborative, Text \rangle$ for behavior-content alignment is formulated in Eq.\ref{eq:cl_bt}. The adaptive alignment strength $w$ is employed as the weight of $\mathcal{L}_{align_{b,v}}$. Similary, the contrastive learning of $\langle Collaborative, Visual \rangle$ is formulated in Eq.\ref{eq:cl_bv}, where $\tau$ is the temperature coefficient, and $\tau=0.07$ in our experiment, $\mathrm{sim}(\cdot, \cdot)$ is the cosine similarity.
\begin{gather}
\label{eq:cl_bt}
 \mathcal{L}_{align_{b,t}} = - \log \frac{\exp(\text{sim}(h_t, h_{b^{+}}) / \tau)}{\exp(\text{sim}(h_t, h_{b^{+}}) / \tau) + \sum_{i=1}^{B-1} \exp(\text{sim}(h_t, h_{b_{i}^{-}}) / \tau)}\\
\label{eq:cl_bv} 
    \mathcal{L}_{align_{b,v}} = - \log \frac{\exp(\text{sim}(h_v, h_{b^{+}}) / \tau)}{\exp(\text{sim}(h_v, h_{b^{+}}) / \tau) + \sum_{i=1}^{B-1} \exp(\text{sim}(h_v, h_{b_{i}^{-}}) / \tau)}
\end{gather}
In addition, we align the text and visual modalities. Unlike the asymmetric alignment from content to behavioral representations, this content-to-content alignment is symmetric, as formalized in Eq.~\ref{eq:content_cl}. The total alignment loss is formalized in Eq.~\ref{eq:cl_total}.
\begin{gather}
     \begin{split}
    \label{eq:content_cl} 
        \mathcal{L}_{align_{v,t}} = -\log \frac{\exp(\text{sim}(h_t, h_{v^{+}}) / \tau)}{\exp(\text{sim}(h_t, h_{v^{+}}) / \tau) + \sum_{i=1}^{B-1} \exp(\text{sim}(h_t, h_{v_{i}^{-}}) / \tau)}\\ -\log \frac{\exp(\text{sim}(h_v, h_{t^{+}}) / \tau)}{\exp(\text{sim}(h_v, h_{t^{+}}) / \tau) + \sum_{i=1}^{B-1} \exp(\text{sim}(h_v, h_{t_{i}^{-}}) / \tau)}\\
    \end{split} \\
    \label{eq:cl_total}
    \mathcal{L}_{align\_total}= \mathcal{L}_{align_{v,t}} +w \mathcal{L}_{align_{b,v}} +w \mathcal{L}_{align_{b,t}} 
\end{gather}

\subsection{Dynamic Behavioral Weighting Mechanism}
\label{sec:section3_4}
To address the challenge that informative SIDs are easily overwhelmed by noisy SIDs under the equal-weight SID paradigm, we propose a dynamic behavioral weighting mechanism that adaptively learns importance scores for behavioral SIDs, enabling the downstream recommendation to suppress noisy SIDs and improve the recommendation accuracy.

\subsubsection{Dynamic Behavioral Weighting Gate}
The dynamic behavioral weighting gate is proposed to learn importance scores for behavioral SIDs based on the L2-magnitude of the pre-trained collaborative embedding. The formulation is in Eq.\ref{eq:router}.
\begin{equation}
\label{eq:router}
R(\mathbf{e_b})=\sigma(N_{\text{norm}}(\mathbf{e}_{b}))* relu(MLP(\mathbf{e}_{b})+b)\\    
\end{equation}
Here, the $MLP$ extracts behavior‑specific semantics. $\sigma(\cdot)$ denotes the sigmoid function, which rescales the learned weights.
This gate suppresses noisy behavioral SIDs, improving the robustness of behavioral SIDs when incorporated with downstream recommendations.

\subsubsection{Sparsely-Activated Training Strategy}
The disparity in collaborative signals between head and long-tail items leads to low importance scores for most SIDs, which may result in insufficient training of the behavioral experts as well as imbalanced training across experts. Inspired by ReMoE\cite{ReMoE}, we adopt a load-balancing strategy with $relu$-based router that effectively alleviates training imbalance among behavioral experts, as formulated in Eq.\ref{eq:reg_i}. $f_{lb}$ is a load-balancing function and $\lambda_i$ is a learnable parameter, which follows ReMoE\cite{ReMoE}.
\begin{gather} 
\label{eq:reg_i}
L_{reg_i}=\lambda_i\frac{1}{B}\sum_t^{B}\sum_j^{N_b} f_{lb}(||R(\mathbf{e_b})_j||)
\end{gather}
The regularization loss $L_{reg}$ then penalizes the deviation from the load-balancing target.

The dynamic behavioral weighting gate is trained with the load-balancing strategy.
This training strategy allows long-tail items to activate only a subset of both behavioral experts and behavioral SIDs, thereby reducing the impact of collaborative noise while ensuring that each expert is sufficiently trained

%% file: secs/exp_v3.tex
In this paper, we conduct extensive experiments on both industrial and public datasets to evaluate the effectiveness of our proposed framework and address the following questions: 
\begin{itemize}[leftmargin=*]
\item \textbf{RQ1}: How does ADC-SID compare to state-of-the-art(SOTA) item tokenizers in terms of quantization metrics and downstream performance in generative retrieval and discriminative ranking tasks?
\item \textbf{RQ2}: What is the contribution of each component in ADC-SID? 
\item \textbf{RQ3}: How sensitive is ADC-SID's performance to its key hyperparameters?
\item \textbf{RQ4}: How effective is our proposed ADC-SID in improving recommendations for items with varying degrees of popularity, especially for those in the long-tail items?
\end{itemize}
\begin{table}[htbp]
    \centering
    \caption{Statistics of Industrial and Public Datasets.}
    \renewcommand{\arraystretch}{1.1}
    \label{fig:data_scale_comparison}
    \resizebox{180pt}{!}{  
    \begin{tabular}{c|c|c}
        \toprule
        Dataset  & Industrial Dataset &Beauty\\
        \midrule
        \#User &35,154,135& 22,363   \\
        \#Item & 48,106,880& 12,101   \\
        \#Interaction&75,730,321,793 & 198,360  \\
        \bottomrule
    \end{tabular}
    }
\end{table}

\begin{table*}[htbp]
\centering
\caption{Overall performance comparison on two datasets. We evaluate all methods on two downstream tasks: generative retrieval and discriminative ranking. Best results in each column are in \textbf{bold}. Our model, ADC-SID, is highlighted in gray. The last row (Improv.) denotes the relative improvement of ADC-SID over the best baseline. The best baseline performance score is denoted in \underline{underline}.}
\label{tab:main_comparison}
\begin{subtable}{\textwidth}
    \centering
    \caption{Generative Retrieval Evaluation}
    \label{tab:generative_results}
    \resizebox{\textwidth}{!}{%
    \begin{tabular}{l | ccc|cccc | ccc|cccc}
        \toprule
        \multirow{2}{*}{\textbf{Methods}} & \multicolumn{7}{c|}{\textbf{Industrial Dataset}} & \multicolumn{7}{c}{\textbf{Amazon Beauty}} \\
        \cmidrule(lr){2-8} \cmidrule(lr){9-15}
        & $L_{\text{recon}}\downarrow$ & Entropy$\uparrow$ & Util.$\uparrow$ & R@50$\uparrow$ & R@100$\uparrow$ & N@50$\uparrow$ & N@100$\uparrow$ & $L_{\text{recon}}\downarrow$ & Entropy$\uparrow$ & Util.$\uparrow$ & R@50$\uparrow$ & R@100$\uparrow$ & N@50$\uparrow$ & N@100$\uparrow$ \\
        \midrule
        RQ-VAE &\underline{0.0033} & 4.2481 & \underline{1.0000} & 0.1854 & 0.2083 &0.1337  &0.1421  &0.6028 & 3.4904 & 0.9900 &0.1213  &0.2398  & 0.0803  &0.1304  \\
        OPQ &0.0038 &4.3984 &1.0000 &0.1872& 0.2105 &0.1432  & 0.1508 &0.9647 & 3.3980 & 0.9600&0.1117  &0.2189  & 0.0802 & 0.1302   \\
        RQ-Kmeans &0.0065 &\underline{4.7232} &\underline{1.0000}&0.1874  &0.2203  &0.1466  &0.1578&0.6834 & 3.7300 & \underline{1.0000}& 0.1283 &0.2285  &0.0840& 0.1347\\
        LETTER &0.0054 &4.2072 & \underline{1.0000} & 0.1882 & 0.2213 &0.1582  & 0.1675 &0.5431 & 2.6819 & \underline{1.0000}& 0.1513  &0.2492 & 0.0937 & 0.1453 \\
        DAS &0.0051 &4.3539 &\underline{1.0000}&0.1884  &0.2238  &0.1586  &0.1697  &0.5432 & 3.6819 & \underline{1.0000}&  0.1503&0.2403  &0.0933  &0.1445  \\
        MM-RQ-VAE &0.0055  &4.2125 &0.9850&\underline{0.2181} & \underline{0.2542} & \underline{0.1592}& \underline{0.1707}
        & \underline{0.5079}& \underline{3.8242}  &\underline{1.0000}& \underline{0.1673} & \underline{0.2598} & \underline{0.0945} & \underline{0.1461} \\
       \midrule
        \rowcolor{gray!15}
        \textbf{ADC-SID(Ours)} & \textbf{0.0031} & \textbf{5.0975} & \textbf{1.0000} & \textbf{0.2774} & \textbf{0.2927} & \textbf{0.1688} & \textbf{0.1744} &  \textbf{0.4470} & \textbf{4.4206} & \textbf{1.0000} & \textbf{0.1855} & \textbf{0.2885} & \textbf{0.0996} & \textbf{0.1568}\\
        \textit{Improv.} & +6.06\%&+7.92\% & +0.00\%& +27.19\% & +15.15\% &+6.03\%  & +2.17\% & +11.99\%  & +15.60\% & +0.00\%&+10.87\% &+11.10\%&+5.40\%  &+7.32\%  \\
        \bottomrule
    \end{tabular}
    } 
\end{subtable}


\vspace{1em} %
\begin{subtable}{\textwidth}
    \centering 
    \caption{Discriminative Ranking Evaluation}
    \label{tab:discriminative_results}
    \footnotesize 
    \begin{tabular}{l | ccc|cc | ccc|cc}
        \toprule
        \multirow{2}{*}{\textbf{Methods}} & \multicolumn{5}{c|}{\textbf{Industrial Dataset}} & \multicolumn{5}{c}{\textbf{Amazon Beauty}} \\
        \cmidrule(lr){2-6} \cmidrule(lr){7-11}
        & $L_{\text{recon}}\downarrow$ & Entropy$\uparrow$ & Util.$\uparrow$ & AUC$\uparrow$ & GAUC$\uparrow$ & $L_{\text{recon}}\downarrow$ & Entropy$\uparrow$ & Util.$\uparrow$ & AUC$\uparrow$ & GAUC$\uparrow$ \\
        \midrule
        RQ-VAE & \underline{0.0033} & 4.2481 &\underline{ 1.0000} & 0.7071 & 0.5805 &0.6028 & 3.4904 & 0.9900 & 0.6446 & 0.5852\\
        OPQ  & 0,0038 &4.3984 &1.0000  &0.7088 & 0.5830 & 0.9647 & 3.3980 & 0.9600&0.6449 &0.5898 \\
        
        RQ-Kmeans  &0.0065  & \underline{4.7232} &\underline{1.0000}  &0.7089  & 0.5832 &0.6834 & 3.7300 & \underline{1.0000}&0.6452 & 0.5909\\
        LETTER  &0.0054  &4.2072 & \underline{1.0000} & 0.7089  & 0.5828 & 0.5431 & 2.6819 & \underline{1.0000} &0.6444  & 0.5973 \\
        DAS  & 0.0051 &4.3539 &\underline{1.0000}  & 0.7091 & \underline{0.5845} &0.5432 & 3.6819 & \underline{1.0000}& \underline{0.6466} & 0.5933 \\
        MM-RQ-VAE  & 0.0055&4.2125 &0.9850  &\underline{0.7095}  & 0.5843 &  \underline{0.5079}&\underline{3.8242} &1.0000  &0.6453  &  \underline{0.5991}\\
        \midrule
       \rowcolor{gray!15}
        \textbf{ADC-SID(Ours)} & \textbf{0.0031} & \textbf{5.0975} & \textbf{1.0000} & \textbf{0.7101} & \textbf{0.5846} & \textbf{0.4470} & \textbf{4.4206} &\textbf{1.0000}  &  \textbf{0.6480}&\textbf{0.6125}  \\
        \textit{Improv.} &+3.03\%&+7.92\% & +0.00\% & +0.07\% & +0.02\% &+11.99\%  & +15.60\% & +0.00\% &+0.12\%  & +2.10\% \\
        \bottomrule
    \end{tabular}
\end{subtable}


\end{table*}

\subsection{Experimental Setup}
\subsubsection{Dataset} We evaluate the proposed framework both on an industrial dataset and a public dataset. 

\textbf{Industrial Dataset}: 
This dataset was collected from a leading e-commerce advertising platform in Southeast Asia between October 2024 and May 2025, encompassing 30 million users and 40 million advertisements. It contains user behavior sequences with an average length of 128, supplemented by rich, multimodal item attributes such as images, titles, and descriptions. Its scale and complexity make it an ideal benchmark for evaluating performance in real-world scenarios.
\begin{table*}
\centering
\captionsetup{width=0.9\textwidth, justification=centering} 
\caption{Ablation Experiments.}
\label{tab:abla_performance}
\begin{tabular}{l|ccc|cccc|cc}
\toprule
Variants&$\mathbf{L}_{recon}\downarrow$ & Entropy$\uparrow$ & Util.$\uparrow$
 & R@50$\uparrow$ & R@100$\uparrow$&N@50$\uparrow$ & N@100$\uparrow$ &AUC$\uparrow$ &GAUC$\uparrow$
\\ 
\midrule
\textbf{ADC-SID} & \textbf{0.0031} &\textbf{ 5.0975} & \textbf{1.0000 }& \textbf{0.2774} & \textbf{0.2927} & \textbf{0.1688} &\textbf{ 0.1744} & \textbf{0.7101} &\textbf{0.5846}\\
w/o Alignment Strength Controller  & 0.0032 & 5.0710 & 1.0000 & 0.2701 & 0.2854 & 0.1618 & 0.1643 & 0.7104 & 0.5845\\
w/o Behavior-Content Contrastive Learning   & 0.0032 & 5.1153 & 1.0000 & 0.2733 & 0.2874 & 0.1653 & 0.1676 & 0.7097 & 0.5846\\
w/o Dynamic Behavioral Weighting Gate  & 0.0033 & 5.0896 & 1.0000 & 0.2705 & 0.2861 & 0.1616 & 0.1641 & 0.7098 & 0.5845\\
w/o Sparsely-Activated Training Strategy& 0.0034 & 5.0571 & 1.0000 & 0.2757 & 0.2903 & 0.1675 & 0.1698 & 0.7097 & 0.5846\\
\bottomrule

\end{tabular}
\end{table*}

\textbf{Public Dataset}: We conduct experiments on the "Beauty" subset of the Amazon Product Reviews dataset \cite{He_2016} to evaluate our approach. The dataset comprises user review data from May 1996 to September 2014, and the statistic is shown in Table \ref{fig:data_scale_comparison}. For the evaluation in generative retrieval, we apply the 5-core filter to exclude unpopular users and items with fewer than five interaction records. Then, we construct user behavior sequences according to chronological order and uniformly set the maximum item sequence length to 20. For the evaluation in discriminative ranking, we use the user-item rating dataset from Amazon Product Reviews. Following standard practice, we treat ratings greater than 3 as positive labels and those less than or equal to 3 as negative labels. The dataset is sorted chronologically, and the first 90\% of the data are used for training, while the remaining 10\% constitute the test set.


\subsubsection{Evaluation Metrics}We evaluate the effectiveness of proposed ADC-SID from both quantization metrics and recommendation metrics.

\textbf{Quantization Metrics}
\begin{itemize}
\item \textbf{Reconstruction Loss} \cite{deng2025onerecunifyingretrieverank} is utilized to evaluate the reconstrction fidelity for the original input vector. 
\item \textbf{Token Distribution Entropy} \cite{bentz2016wordentropynaturallanguages} is utilized to evaluate the diversity of the distribution across semantic codewords in codebooks.
\item \textbf{Codebook utilization} \cite{zhu2024scalingcodebooksizevqgan} is employed to reflect the efficiency with which the model uses the codebook vectors. 
\end{itemize}

\textbf{Recommendation Metrics}: In generative retrieval, Recall@N and NDCG@N with N=50, 100 are used to evaluate the performance. In discriminative ranking, AUC and GAUC are used to evaluate the prediction accuracy. GAUC is the most important metric for our personalized ads system. For the online experiments, the Advertising Revenue, Click-Through Rate (CTR) are used to evaluate the online performace.

\textbf{Baselines}
We compare our proposed method with SOTA Semantic ID generation approaches, which can be categorized into two groups: (1) Content-based SIDs (e.g., RQ-VAE, OPQ), (2) Behavior-content aligned SIDs (e.g., RQ-Kmeans, DAS\cite{ye2025dasdualalignedsemanticids}, LETTER, MM-RQ-VAE). 
\begin{itemize}
\item \textbf{RQ-VAE\cite{zhu2024scalingcodebooksizevqgan}}: TIGER\cite{rajput2023recommendersystemsgenerativeretrieval} transforms content features such as titles, item descriptions, and categories into textual embeddings using a pre-trained LLM. It then employs RQ-VAE to quantize these embeddings into hierarchical SIDs.
\item \textbf{OPQ\cite{ge_optimized_nodate}}: RPG\cite{hou2025generatinglongsemanticids} introduce Optimized Product Quantization(OPQ) to convert pretrained textual embeddings into a tuple of unordered SIDs.
\item \textbf{RQ-Kmeans\cite{luo2024qarmquantitativealignmentmultimodal}}: One-rec\cite{zhou2025onerectechnicalreport} integrates RQ-VAE and K-means to quantize behavior-finetuned multimodal representations of items in a coarse-to-fine manner, where K-means clustering is applied to the residuals.
\item \textbf{DAS\cite{ye2025dasdualalignedsemanticids}}: DAS introduces multi-view contrastive alignment to maximize mutual information between SIDs and collaborative signals. This process generates hierarchical, behavior-aware content SIDs using RQ-VAE.
\item \textbf{LETTER\cite{wang2024learnableitemtokenizationgenerative}}: LETTER integrates hierarchical semantics, collaborative signals, and code assignment diversity to generate behavior-content-fused SIDs using RQ-VAE.
\item \textbf{MM-RQ-VAE\cite{wang2025empoweringlargelanguagemodel}}: MM-RQ-VAE generates collaborative, textual, and visual SIDs from pre-trained collaborative and multimodal embeddings. It also introduces contrastive learning for behavior-content alignment.
\end{itemize}

\subsubsection{Implementation Details} In this section, we introduce the implementation details.

\textbf{Recommendation Foundations}: For generative retrieval tasks, we adopt an encoder-decoder generative recommendation framework, REG4REC \cite{xing2025reg4recreasoningenhancedgenerativemodel} to evaluate the performance of different SIDs. 
For discriminative ranking task, we employ the well-established Parameter Personalized Network (PPNet) \cite{chang2023pepnetparameterembeddingpersonalized} as the backbone architecture.

\textbf{Codebook Setting}: In the industrial dataset, the codebook size is set to 3,00 and the length of SIDs is set to 10 for ADC-SID and baselines. Specifically, $N_s=2$, $N_t=2$, $N_v=2$, $N_b=4$ are set for ADC-SID. In public datasets, the codebook size is set to 100, the length is set to 6. Specifically, $N_s=1$, $N_t=1$ and $N_v=1$,$N_b=3$ are set for ADC-SID. The pre-trained collaborative representations are obtained from the SASRec model~\cite{kang2018selfattentivesequentialrecommendation}, which is trained on user-item interactions without L2 normalization during inference. The pre-trained text representations are obtained from Qwen3-Embedding 7B \cite{zhang2025qwen3embeddingadvancingtext}. The pre-trained visual representations are obtained from PailiTAO v8, which is trained on data from an e-commerce advertising platform in Asia.

\textbf{Dynamic-Weight Behavioral SID Incorporate with Downstream Recommendations}
To address the challenge that crucial signals are easily overwhelmed by noisy SIDs under the equal-weight SID paradigm, the importance weights learned from the dynamic behavioral weighting mechanism are incorporated into downstream recommendation to suppress noisy SIDs. When applied to discriminative ranking, both the mapping between item IDs and SIDs and the learned importance weights are utilized as inputs to the ranking model. In the case of generative retrieval, the learned weights are utilized as loss weights for SID generation.

\begin{table*}
\centering
\captionsetup{width=0.9\textwidth, justification=centering} 
\caption{Hyper-Parameter Analysis on Contrastive Loss Weight}
\label{tab:cl_performance}
\begin{tabular}{l|ccc|cccc|cc}
\toprule
Variants&$\mathbf{L}_{recon}\downarrow$ & Entropy$\uparrow$ & Utilization$\uparrow$
 & R@50$\uparrow$ & R@100$\uparrow$&N@50$\uparrow$ & N@100$\uparrow$ &AUC$\uparrow$ &GAUC$\uparrow$
\\ 
\midrule
$\alpha$=20, $\beta$=7  & 0.0032 & 5.0844 & 1.0000 & 0.2749 & 0.2894 & 0.1664 & 0.1688 & 0.7106 & 0.5840\\
$\alpha$=20, $\beta$=9 & 0.0032 & 5.0711 & 1.0000 & 0.2750 & 0.2889 & 0.1677 & 0.1709 & 0.7105 & 0.5842\\
$\alpha$=20, $\beta$=14  & 0.0033 & 5.0967 & 1.0000 & 0.2760 & 0.2911 &0.1686& 0.1707 & 0.7105 & 0.5839\\
\textbf{$\alpha$=10, $\beta$=9} & \textbf{0.0031} & \textbf{5.0975} & \textbf{1.0000} & \textbf{0.2774} & \textbf{0.2927} & \textbf{0.1688} & \textbf{0.1744} & \textbf{0.7101} & \textbf{0.5846}\\
\bottomrule
\end{tabular}
\end{table*}
\begin{figure*}[t]
\includegraphics[width=1\textwidth]{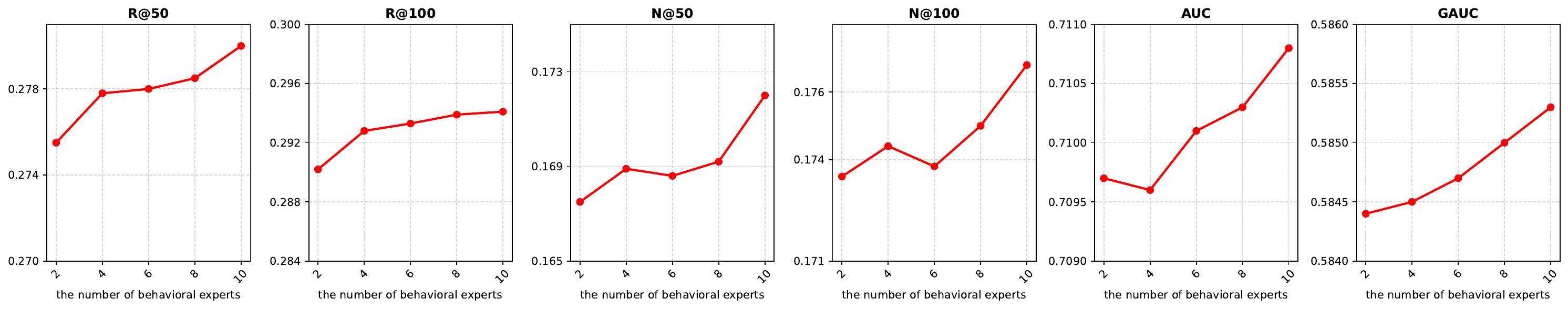}
\centering
\caption{
Hyper-Parameter Analysis on the Number of Behavioral Experts.
}
\label{fig:Sparsity}
\end{figure*}

\subsection{Overall Performance (RQ1)}
We compare our proposed method with state-of-the-art(SOTA) semantic ID approaches for generative retrieval and discriminative ranking tasks on both public and industrial datasets. 
The overall performance comparison is summarized in Table \ref{tab:main_comparison}, the results lead to several key observations: Firstly, behavior-content aligned SIDs (e.g., RQ-Kmeans, LETTER) consistently outperform content-only SIDs (RQ-VAE, OPQ) across R@100, N@100, and AUC. This highlights the fundamental limitation of relying solely on static content and underscores the necessity of incorporating behavioral signals. Secondly, the significant performance lift of MM-RQ-VAE over the original RQ-VAE directly demonstrates the critical value of incorporating collaborative information into the SID generation process. 
Furthermore, the performance comparsion between MM-RQ-VAE and LETTER, DAS, RQ-Kmeans demonstrated that explicitly generating behavioral SIDs is more effective at capturing an item's collaborative patterns, with a better performance of downstream recommendations.


ADC-SID demonstrates superior performance across both generative retrieval and discriminative ranking, outperforming all baselines, including content-only and existing behavior-content aligned SIDs. This superiority stems from its unique design: unlike methods that perform indiscriminate alignment, ADC-SID adaptively fuses the content and collaborative information and adaptively suppresses noise, resulting in a more robust and expressive SID representation. As shown in Table \ref{tab:main_comparison}, ADC-SID achieves substantial gains in Recall@50 and Recall@100, while yielding only modest improvements in NDCG@50 and NDCG@100. This indicates that incorporating richer semantic IDs enables the model to retrieve more items of user interest; however, these retrieved items often receive relatively low scores, leading to limited gains in NDCG. This observation further reflects that a large number of long-tail items suffer from insufficient training, resulting in underestimated relevance scores. Consequently, improving the representation quality of SIDs for long-tail items is crucial.

\subsection{Ablation Study on Industrial Dataset (RQ2)}
We conduct the ablation experiments in Table \ref{tab:abla_performance} to study how each module contributes to the overall performance of ADC-SID.

\subsubsection{Impact of Adaptive Behavior-Content Alignment}
\textbf{w/o the Alignment Strength Controller}: In this experiment, we disable the alignment strength controller and align content representations with collaborative embeddings indiscriminately. The performance degradation has demonstrated that it's significant to mitigate the collaborative noise from long-tail items that corrupts behavior–content alignment.

\textbf{w/o the Behavior-Content Contrastive Learning}: In this experiment, we disable the adaptive behavior-content Alignment module, which leads to a consistent drop in both Recall and NDCG across all settings. This finding indicates a substantial modality gap between the content and behavioral domains, which hinders the model's ability to learn the shared information. The resulting performance degradation is therefore expected, as the contrastive learning component is crucial for bridging this gap and enabling effective behavior-content information fusion.

\subsubsection{Impact of Dynamic Behavioral Weighting Mechanism}
\textbf{w/o the Dynamic Behavioral Weighting Gate}: Removing the dynamic behavioral weighting gate impairs the model’s ability to be aware of the importance of behavioral SIDs, leading to a drop in recommendation accuracy on both discriminative ranking and generative retrieval tasks. This demonstrates that it's is crucial to enable the downstream recommendation to suppress noisy SIDs for improving downstream recommendation performance

\textbf{w/o Sparsely-Activated Training Strategy}:
Removing the sparsely-activated training strategy also leads to performance degradation. The sparsely-activated training strategy plays a significant role in the behavioral experts. This training strategy encourages the activation of every expert and codebook during training, thereby alleviating the load imbalance issue. Therefore, during training, the behavioral experts can adaptively activate a varying number of experts for head and long-tail items, thereby reducing the impact of noise on the quantization network.



\subsection{Hyper-Parameter Analysis(RQ3)}
We further investigate how various hyper-parameter settings affect the model’s performance on industrial dataset.


\textbf{The Impact of Alignment Strength Controller Hyperparameters} 
We conducted a hyperparameter experiment on ($\alpha$, $\beta$) for the Alignment Strength Controller, testing four configurations that yield diverse weighting curves in Fig.\ref{fig:controller}. As shown in Table \ref{tab:cl_performance}, the setting ($\alpha$=10, $\beta$=9) achieved the highest recommendation accuracy. This optimal result suggests that for this dataset's distribution, noise filtering is most effective when applied to approximately the 40\% least frequent (long-tail) items. The tunability of parameters $\alpha$ and $\beta$ underscores the inherent flexibility of our design, allowing it to adapt to diverse data landscapes.

\textbf{The Impact of the Number of Behavioral Experts} 
As the number of behavioral experts increases, the encoder capacity expands, leading to stronger representational abilities. Consequently, a significant increase in recommendation accuracy is observed across both discriminative ranking and generative retrieval tasks, as shown in Fig.\ref{fig:Sparsity}. Besides, ADC-SID demonstrates an advantage in dynamic-weight behavioral SIDs: under the equal-weight behavioral SID paradigm, when the number of semantic IDs is increased, the number of behavioral semantic IDs for long-tail items also grows proportionally, introducing more long-tail noise into more semantic IDs. The downstream recommendation task struggles to distinguish noisy SIDs from informative ones. In contrast, our proposed dynamic-weight behavioral SIDs learn importance scores for semantic IDs, enabling better incorporation with the downstream task and effectively suppressing noisy SIDs.

\subsection{Item Popularity Stratified Performance Analysis (RQ4)}
\begin{figure}[t]
  \centering %
  \includegraphics[width=1\columnwidth]{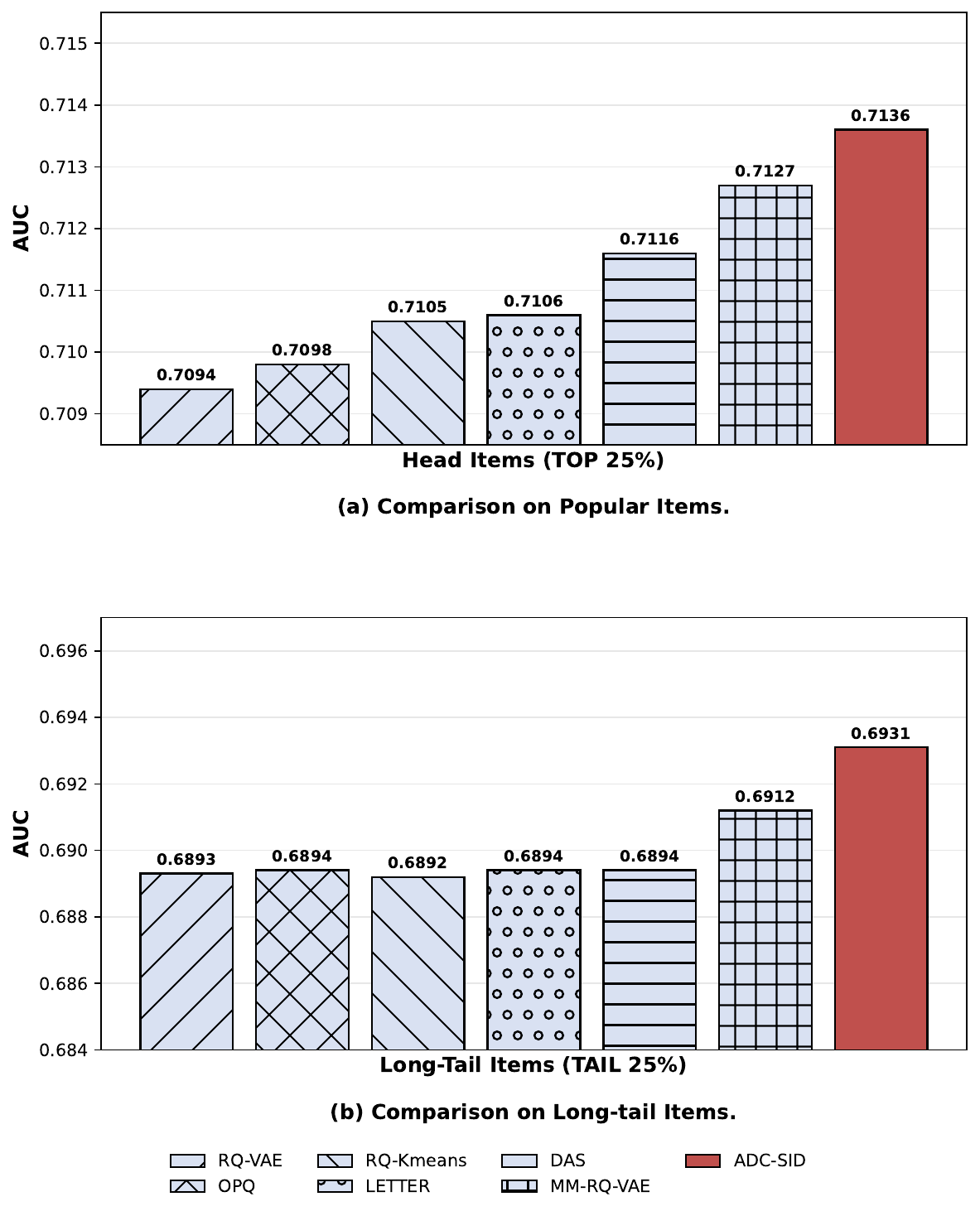}
  
  \caption{Performance comparison stratified by item popularity on popular and long-tail items.}
  \label{fig:popularity_stratification} %
  \vspace{-0.3cm}
\end{figure}

To investigate the performance of different item popularity stratification, we perform a stratified analysis based on item popularity. We categorize items into 'popular' (top 25\%) and 'long-tail' (bottom 25\%) groups based on their impression counts over the last 30 days, and then evaluate the AUC for each group, as shown in Fig.\ref{fig:popularity_stratification}.

For head items, content-based SID performs worse than other SID quantization methods that incorporate collaborative information in ranking tasks. Integrating collaborative information is crucial, as it enhances the SIDs' expressiveness to capture complex behavioral patterns, thereby improving performance. Furthermore, our proposed ADC-SID advances beyond previous approaches by adaptively denoising collaborative signals for SID quantization. This process yields a significantly more expressive semantic representation.
For tail items: 
ADC-SID achieves the largest performance gain. The adaptive behavior-content alignment shields the stable content representations of tail items from their noisy collaborative signals. Concurrently, the dynamic behavioral weighting mechanism learns importance scores for behavioral SIDs, enabling the downstream recommendation to suppress uninformative SIDs and improve downstream recommendation performance.
This dual mechanism significantly boosts performance on the long tail.

By adaptively balancing the expressive independence of head items with the generalization capability for tail items, our method generates more robust and effective SIDs.